\documentstyle[12pt,epsfig]{article}
\begin{document} 
\font\ninerm = cmr9 
 
\def\footnoterule{\kern-3pt \hrule width \hsize \kern2.5pt}

\pagestyle{empty} 
 
\begin{flushright} 
astro-ph/0008107 \\ 
August 2000 
\end{flushright} 
 
\vskip 0.5 cm

\begin{center} 
{\large\bf Planck-scale deformation of Lorentz symmetry as a\\ 
solution to the UHECR and the TeV-$\gamma$ paradoxes} 
\end{center} 
\vskip 1.5 cm 
\begin{center} 
{\bf Giovanni AMELINO-CAMELIA}$^a$ and {\bf Tsvi PIRAN}$^b$\\ 
\end{center} 
\begin{center} 
{\it $^a$Dipart.~Fisica, 
Univ.~Roma ``La Sapienza'', 
P.le Moro 2, 00185 Roma, Italy}\\ 
{\it $^b$Racah Institute of Physics, Hebrew University, 
Jerusalem 91904, Israel} 
\end{center} 
 
\vspace{1cm} 
\begin{center} 
{\bf ABSTRACT} 
\end{center} 
 
{\leftskip=0.6in \rightskip=0.6in  
   
  One of the most puzzling current experimental physics paradoxes is 
  the arrival on Earth of Ultra High Energy Cosmic Rays (UHECRs) with 
  energies above the GZK threshold ($5 {\times} 10^{19}$eV). 
  Photopion production by CMBR photons should reduce the energy of 
  these protons below this level.  The recent observation of 20TeV 
  photons from Mk 501 (a BL Lac object at a distance of 150Mpc) is 
  another somewhat similar paradox.  These high energy photons should 
  have disappeared due to pair production with IR background photons. 
  A common feature of these two paradoxes is that they can both be 
  seen as ``threshold anomalies": energies corresponding to an 
  expected threshold (pion production or pair creation) are reached 
  but the threshold is not observed.  Several (relatively speculative) 
  models have been proposed for the UHECR paradox. No solution has yet 
  been proposed for the TeV-$\gamma$ paradox. Remarkably, the single 
  drastic assumption of a violation of ordinary Lorentz invariance 
  would resolve both paradoxes.  We present here a formalism for the 
  systematic description of the type of Lorentz-invariance deformation 
  (LID) that could be induced by non-trivial short-distance structure 
  of space-time, and we show that this formalism is well suited for 
  comparison of experimental data with LID predictions.  We use the 
  UHECR and TeV-$\gamma$ data, as well as upper bounds on 
  time-of-flight differences between photons of different energies, to 
  constrain the parameter space of the LID.  A model with only two 
  free parameters, an energy scale and a dimensionless parameter 
  characterizing the functional dependence on the energy scale, is 
  shown to be sufficient to solve both the UHECR and the TeV-$\gamma$ 
  threshold anomalies while satisfying the time-of-flight bounds. The 
  allowed region of the two-parameter space is relatively small, but, 
  remarkably, it fits perfectly the expectations of the 
  quantum-gravity-motivated space-time models known to support such 
  deformations of Lorentz invariance: an integer value of the 
  dimensionless parameter and a characteristic energy scale constrained 
  to a narrow interval in the neighborhood of the Planck scale.} 
  
\newpage 
 
\baselineskip 12pt plus .5pt minus .5pt 

\pagenumbering{arabic} 
\pagestyle{plain}  
 
\section{Introduction} 
Significant evidence has accumulated in recent years suggesting that 
in two different regimes, Ultra High Energy Cosmic Rays (UHECRs) and 
multi-TeV photons, the universe is more transparent than what it was 
expected to be.  UHECRs interact with the Cosmic Microwave Background 
Radiation (CMBR) and produce pions. TeV photons interact with the 
Infra Red (IR) photons and produce electron-positron pairs.  These 
interactions should make observations of UHECRs with $E > 5 {\times} 
10^{19}$eV (the GZK limit)~\cite{GZK} or of gamma-rays with $E > 
20$TeV from distant sources 
unlikely~\cite{Nikishov62,Gould67,Stecker92}.  Still UHECRs above the 
GZK limit and 20TeV photons from Mk 501 are observed. 
 
Numerous solutions have been proposed for the UHECR paradox (see 
\cite{Olinto} for a recent review).  Most of these solutions require 
new Physics. There are practically no proposals concerning the 
TeV-$\gamma$ paradox (see however, \cite{Harwit99}).  It is striking 
that there are some common features in these otherwise apparently 
unrelated paradoxes.  In both cases low energy photons interact with 
high energy particles. The reactions should take place because when 
Lorentz transformed to the CM frame the low energy photon have 
sufficient energy to overcome an intrinsic threshold.  In both cases 
the CM energies are rather modest ($\sim 100 $ MeV for UHECRs and 
$\sim 1$ MeV for the TeV photons) and the physical processes involved 
are extremely well understood and measured in the laboratory.  In both 
cases we observe particles above a seemingly robust threshold and the 
observations can be considered as a ``threshold anomaly''.  It is 
remarkable that in spite of these similarities at present there is 
only one mechanism that could resolve both paradoxes: a mechanism 
based on the single, however drastic, assumption of a violation of 
ordinary Lorentz invariance. 
 
The possibility that the cosmic-ray threshold anomaly could be a 
signal of violation of ordinary Lorentz invariance had already been 
emphasized in Refs.~\cite{Gonzalez,colgla,ita,bertli,sato}.  In this 
work we combine these earlier points with the very recent 
suggestion~\cite{kifu,kluz,Protheroe_Meyer} that Lorentz-invariance 
violation could be the origin of the TeV-$\gamma$ threshold anomaly. 
We analyze a general phenomenological framework for the description of 
the type of Lorentz-invariance deformation (LID) that could be induced 
by non-trivial short-distance structure of space-time, and we ask 
whether there are choices of LID parameters that would simultaneously 
solve the two threshold anomalies while satisfying the constraints 
imposed by the fact that the results of experimental 
searches~\cite{schaef,billetal} of energy-dependent relative delays 
between the times of arrival of simultaneously emitted photons are 
still consistent with ordinary Lorentz invariance.  We obtain, under 
these assumptions, strict limits on the possible parameter space of 
LID.  The fact that one is at all able to give a quantitative 
description of both threshold anomalies with a simple two-parameter 
LID model provides encouragement for the interpretation of the data as 
a sign of LID; moreover, it is quite remarkable that the values 
expected from quantum-gravity considerations (most notably the energy 
scale characterizing the deformation being given by the Planck scale) 
are in agreement with the strict limits we derive. 
 
We review in Sections 2 and 3 the observational background and the 
theoretical problems related to the observations of UHECRs (Section 2) 
and TeV photons (Section 3). In Section 4 we describe a special (two 
parameter) model for LID and we obtain limits on these two parameters. 
In Section 5 we describe a more general five-parameter LID formalism 
and again we constrain the parameter space with the available data. In 
Section 6 we compare our formalism with the Coleman and 
Glashow~\cite{colgla} formalism for Planck-scale-independent Lorentz 
invariance violations.  We summarize our results in section 7. 
Appendix A is devoted to the $\kappa$-Minkowski space-time, which is 
an example of quantum-gravity motivated space-time that allows a 
simple illustration of some of the structures here considered. 
 
\section{UHECRs and the GZK paradox} 
 
The high energy cosmic rays (CR) spectrum depicts a clear break at 
$\sim 5 {\times} 10^{18}$eV. This break is accompanied by a transition 
in the CR composition from nuclei to protons.  Above this break the 
spectrum behaves (with a decreasing statistical certainty due to the 
small number of events) as a single power law $N(E) \sim E^{-2.7}$ all 
the way up to $3.2 {\times} 10^{20}$eV \cite{flyseye}, the highest 
energy CR observed so far. 
 
A sufficiently energetic CMBR photon, at the tail of the black body 
thermal distribution, is seen in the rest frame of an Ultra High 
Energy (UHE) proton with $E >5 {\times} 10^{19}$eV as a $> 140$MeV 
photon, above the threshold for pion production.  UHE protons should 
loose energy due to photopion production and should slow down until 
their energy is below the GZK energy\footnote{The exact composition of 
  UHECRs is unknown and it is possible that UHECRs are heavy nuclei 
  rather than protons. In this case such nuclei would undergo 
  photodisintegration when interacting with CMBR photons.  The 
  threshold energy for a photodisintegration of a nuclei is several 
  MeV. It just happens to be true, purely as a result of a numerical 
  coincidence, that the threshold is reached when the energy of a 
  typical nuclei, say Fe, is $\sim 5 {\times} 10^{19}$eV.  Thus the 
  GZK paradox is insensitive to the question of what is the exact 
  composition of UHECRs.}.  The process stops when CMBR photons 
energetic enough to produce pions are not sufficiently 
abundant~\cite{GZK}.  The proton's mean free path in the CMBR 
decreases exponentially with energy (down to a few Mpc) above the GZK 
limit ($\sim 5 {\times} 10^{19}$eV). Yet more than 15 CRs have been 
observed with nominal energies at or above $10^{20} {\pm} 30\%$ 
eV\cite{AGASA98,watson:HEPiN}. 
 
There are no astrophysical sources capable of accelerating particles 
to such energies within a few tens of Mpc from us (at least not in the 
direction of the observed UHECRs). Furthermore if the CRs are produced 
homogeneously in space and time, we would expect a break in the CR 
spectrum around the GZK threshold: below the threshold we would 
observe CRs from the whole universe; above the threshold we would 
observe CRs only from the nearest few Mpc. The corresponding jump by a 
factor of $\sim 30-100$ in the extrapolated number counts above and 
below the threshold, is not seen. 
 
Numerous solutions have been proposed to resolve the GZK paradox (see 
\cite{Olinto} for a recent review). These solutions include, among 
others, new physics solutions like the decay of topological defects, 
weakly interacting messengers like $S_0$ or neutrinos with anomalous 
cross sections at high energies (the `Z-burst" model). Conventional 
astrophysics solutions like acceleration of UHECRs by GRBs or local 
AGNs require the ad hoc assumption that Earth is located in a not 
generic place in space-time (we should be nearer than average to a 
typical source by a factor of 5) as well as very strong intergalactic 
magnetic fields~\cite{Farrar_Piran2}. Another conventional solution, 
the acceleration of Fe nuclei by magnetars in the galactic halo, 
requires a new, otherwise unobserved, population of galactic halo 
objects. Clearly there is no simple conservative solution to this 
puzzle. 
 
 From the point of view of our LID phenomenology it is important to 
notice that for a solution of the GZK paradox it would be necessary 
(and sufficient) for LID to push the threshold energy upwards by a 
factor of six. In fact, the mean free path of a $5 {\times} 10^{19}$eV 
proton is almost a Gpc, while the highest observed UHECR energy is 
$3.2 {\times} 10^{20}$eV. 
 
\section{TeV photons from Mk 501 and Mk 421} 
 
HEGRA has detected high-energy photons with a spectrum ranging up to 
24 TeV \cite{Aharonian99} from Markarian 501 (Mk 501), a BL Lac 
object at a redshift of 0.034 ($\sim 157$ Mpc). This observation 
indicates a second paradox of a similar nature. A high energy photon 
propagating in the intergalactic space can interact with an IR 
background photon and produce an electron-positron pair if the CM 
energy is above $2m_ec^2$. The maximal wavelength of an IR photon that 
could create a pair with a 10 TeV photon is 40$\mu m$. As the cross 
section for pair creation peaks at a center of mass energy of about $3 
m_e c^2$, 10 TeV photons are most sensitive to 30$\mu$M IR photon and 
the mean free path of these photons depends on the spectrum of the IR 
photons at the $\sim 15-40 \mu$M range.  These wavelengths scale like 
10TeV/$E$ for different energies. 
 
There have been several attempts to model the IR background resulting 
from different cosmological evolutionary models 
\cite{Primack99,Malkan98,Dwek98,Fall}. Recently, new data from DIRBE 
at 2.2$\mu M$ \cite{Wright00}, at 60 and 100 $\mu$M \cite{Finkbeiner} 
and at 140 and 240 $\mu$M \cite{Hauser98}, and from ISOCOM at 15$\mu$M 
\cite{Biviano99} suggest that the IR background is even higher. 
According to these data the flux of IR photons is $\sim 2.5 {\times} 
10^{-5}$erg cm$^{-2}$sec$^{-1}$sr$^{-1}$ around 60-120 $\mu$M and 
falls off by an order of magnitude towards 15 $\mu$M.  This decrease 
is important as it would lead to a much shorter mean free path for 
20TeV photons as compared to the mean free path of 10TeV photons. 
 
It was originally suggested that the expected break, corresponding to 
hard-photon disappearance in the IR background, in the GeV-to-TeV 
spectrum of AGNs could be used to determine the IR background 
spectrum.  This would have been based on searches of a 
distance-dependent break in the spectrum of various AGNs. However, no 
apparent break is seen in the spectrum of MK 501 at $\sim 20$TeV 
range, where the optical depth seems to exceed unity.  Using current 
IR background estimates Coppi and Aharonian \cite{Coppi99} find an 
optical depth of 5 for 20TeV photons from MK 501 (see also 
\cite{Protheroe_Meyer}). This optical depth increases rapidly with 
energy. Thus, photons at these energies are exponentially suppressed, 
unless they somehow evade the pair-production process. 
 
Unlike the GZK paradox only a few solutions have been proposed for the 
TeV-$\gamma$ paradox.  First, it is possible that there is an upturn 
in the intrinsic spectrum emitted by Mk 501. Such an upturn would 
compensate for the exponential suppression at this region.  Clearly 
this is an extremely fine-tuned solution as the expected energy of 
this upturn should somehow be tuned to the energy at which the optical 
depth from MK 501 to Earth is unity. This energy scale is distance 
dependent and it puts us in a very special position relative to the 
source.  It is of course possible that the IR intensity has been 
overestimated. A shift in the energy estimate of HEGRA would also 
explain the paradox.  Finally Harwit, Protheroe and Biermann 
\cite{Harwit99} suggest that multiple TeV photons may be emitted 
coherently by Mk 501 and if they arrive at Earth very close in time 
and space they may be confused with a single photon event with higher 
energy. 
 
With current data, $\sim 10$TeV photons from Mk 501 could reach Earth, 
while $\sim 20$TeV photons are exponentially suppressed.  This happens 
mainly because of the rapid fall off of the IR spectrum below 
60$\mu$m.  We conclude that a LID upwards shift of the threshold 
energy by a factor of two would resolve this paradox. 
 
Having discussed the relevance of Mk 501 for the emergence of the 
TeV-$\gamma$ threshold anomaly we turn now to TeV photons from Mk 421 
(another BL Lac object at a redshift of 0.031, corresponding 
to $\sim 143$Mpc). 
It is not clear if the spectrum of this source extends high 
enough to pose a paradox comparable to the one indicated by Mk 501. 
However, we note here the simultaneous (within the experimental 
sensitivity) arrival of 1TeV photons and 2TeV from this source.  This 
was used to limit the time-of-flight differences between photons of 
different energies to less than 200 seconds.  This in turn allowed to 
establish, through an analysis of the type proposed in 
Ref.~\cite{grbgac}, an upper limit on Planck-scale-induced 
LID~\cite{billetal} which will be a key element of our 
analysis.  We call these constraints in the following time-of-flight 
constraints. 
 
\section{Lorentz-invariance-violating dispersion relation} 
 
We start by considering first, a class of dispersion relations 
(following~\cite{aemn1,gackpoinplb,grbgac} for $\alpha=1$, 
and~\cite{polonpap,AmePiran00} for a general $\alpha$) which in the 
high-energy regime takes the form: 
\begin{equation} 
E^2 - \vec{p}^2 - m^2 \simeq  \eta E^2 \left({E \over 
    E_{p}}\right)^\alpha 
\simeq  \eta \vec{p}^2 \left({E \over 
    E_{p}}\right)^\alpha 
~. 
\label{dispone} 
\end{equation} 
$m$, $E$ and $\vec{p}$ denote the mass, the energy and the 
(3-component) momentum of the particle, $E_{p}$ is the Planck energy 
scale ($E_{p} \sim 10^{22}$MeV), while $\alpha$ and $\eta$ are free 
parameters characterizing the deviation from ordinary Lorentz 
invariance (in particular, $\alpha$ specifies how strongly the 
magnitude of the deformation is suppressed by $E_{p}$).  Clearly, in 
(\ref{dispone}) the ``speed-of-light constant" $c$ has been set to 
one. (Note however that in this framework $c$ is to be understood as 
the speed of low-energy massless particles~\cite{grbgac}.)  Also 
notice that in (\ref{dispone}) we wrote the deformation term in two 
ways, as a $E^2 (E /E_{p})^\alpha$ correction and as 
a $p^2 (E/E_{p})^\alpha$ correction, 
which are equivalent within our present 
analysis based exclusively on high-energy data, for which $E \sim p$, 
but would be different when studied with respect to low-energy data. 
(Of course, a given short-distance picture of space-time will have 
only one dispersion relation; for example, in ``$\kappa$-Minkowski 
space-time", the space-time which we describe in Appendix~A in order 
to illustrate in an explicit framework some of the structures relevant 
for our analysis, one encounters a deformation of 
type $p^2 (E/E_{p})^\alpha$.) 
 
In previous works~[31-35,9]
a slightly different notation had been used to describe this same class 
of deformations, which in particular replaced our $\eta$ by two 
quantities: the scale $E_{QG} \equiv |\eta|^{-1/\alpha} E_{p}$ and a 
sign variable $\xi_{\pm} \equiv \eta/|\eta|$. 
The $\alpha$,$\eta$ notation turns out to be more suitable for the 
description of the technical aspects of the analysis discussed here, 
but it is useful to keep in mind that the scale of Lorentz-deformation 
is obtained as $|\eta|^{-1/\alpha} E_{p}$. 
 
As hinted by the presence of the Planck scale, our interest in 
deformed dispersion relations of type (\ref{dispone}) originates from 
the fact that such deformations have independently emerged in theory 
work on quantum properties of space-time. We postpone the discussion 
of this motivation to the next Section, where we also clarify 
which types of generalizations of (\ref{dispone}) 
could also be motivated by Planck-scale physics. 
 
While our analysis is motivated by the role that the deformed 
dispersion relation (\ref{dispone}) might have in quantum gravity, one 
could of course consider (\ref{dispone}) quite independently of 
quantum gravity\footnote{Having mentioned that of course the 
  deformation (\ref{dispone}) could be considered independently of its 
  quantum-gravity motivation, let us also mention in passing that even 
  outside the quantum-gravity literature there is a large amount of 
  work on the theory and phenomenology of violations of Lorentz 
  invariance (see, {\it e.g.}, the recent 
  Refs.~\cite{Gonzalez,colgla,carjack,jackost}, which also provide a 
  good starting point for a literature search backward in time).}. The 
quantum-gravity intuition would then be seen as a way to develop a 
theoretical prejudice for plausible values of $\alpha$ and $\eta$. In 
particular, corrections going like $(E/E_{p})^\alpha$ typically emerge 
in quantum gravity as leading-order pieces of some more complicated 
analytic structures~\cite{grbgac,gackpoinplb,kpoinap,gacmaj}.  
This provides, of course, a special motivation for the study 
of the cases $\alpha=1$ 
and $\alpha=2$. [$f(E/E_{p}) \simeq 1 + a_1 (E/E_{p})^{n_1} +...$.]  
Moreover, the fact that in quantum gravity 
the scale $E_{QG}$ is expected to be somewhere between the GUT scale 
and the Planck scale corresponds to the expectation that $\eta$ should 
not be far from the range $1 \le \eta \le 10^{3 \alpha}$. 
 
\subsection{Deformed thresholds from deformed  dispersion relations} 
 
We intend to discuss the implications of Eq.~(\ref{dispone}) for the 
evaluation of threshold momenta.  Before doing that let us briefly 
summarize the derivation of the equation describing the threshold in 
the ordinary Lorentz-invariant case.  Relevant for our 
phenomenological considerations is the process in which the head-on 
collision between a soft photon of energy $\epsilon$ and momentum $q$ 
and a high-energy particle of energy $E_1$ and momentum $\vec{p}_1$ 
leads to the production of two particles with energies $E_2$,$E_3$ and 
momenta $\vec{p}_2$,$\vec{p}_3$.  At threshold (no energy available
for transverse momenta), energy conservation 
and momentum conservation imply 
\begin{equation} 
E_1+\epsilon=E_2+E_3 
~, 
\label{econsv} 
\end{equation} 
\begin{equation} 
p_1-q=p_2+p_3~; 
\label{pconsv} 
\end{equation} 
moreover, using the ordinary 
Lorentz-invariant relation between energy and momentum, 
one also has the relations 
\begin{equation} 
q=\epsilon~,~~~E_i = \sqrt{p_i^2+m_i^2} 
\simeq p_i + {m_i^2 \over 2 p_i} 
~, 
\label{lirel} 
\end{equation} 
where $m_i$ denotes the mass of the particle with momentum $p_i$ and 
the fact that $p_1$ (and, as a consequence, $p_2$ and $p_3$) is a 
large momentum has been used to approximate the square root. 
 
The threshold conditions are usually identified by transforming this 
laboratory-frame relations into CM-frame relations and imposing that 
the CM energy be equal to $m_2+m_3$; however, in preparation for the 
discussion of deformations of Lorentz invariance it is useful to work 
fully in the context of the laboratory frame. There the threshold 
value $p_{1,th}$ of the momentum $p_1$ can be identified with the 
requirement that the solutions for $p_2$ and $p_3$ as a function of 
$p_1$ (with a given value of $\epsilon$) that follow from 
Eqs.~(\ref{econsv}), (\ref{pconsv}) and (\ref{lirel}) should be 
imaginary for $p_1 < p_{1,th}$ and should be real  
for $p_1 \ge p_{1,th}$.   
This straightforwardly leads to the threshold equation 
\begin{equation} 
p_{1,th} \simeq {(m_2 + m_3)^2 - m_1^2 \over 4 \epsilon} 
~. 
\label{lithresh} 
\end{equation} 
 
This standard Lorentz-invariant analysis is 
modified~\cite{ita,sato,kifu,kluz,Protheroe_Meyer,AmePiran00} by the 
deformations codified in (\ref{dispone}).  The key point is that 
Eq.~(\ref{lirel}) should be replaced by 
\begin{equation} 
\epsilon= q + \eta {q^{1+\alpha} \over 2 E_p^\alpha}  
~,~~~ 
E_i \simeq p_i + {m_i^2 \over 2 p_i} 
+ \eta {p_i^{1+\alpha} \over 2 E_p^\alpha} 
~. 
\label{lv1rel} 
\end{equation} 
Combining (\ref{econsv}), (\ref{pconsv}) and (\ref{lv1rel}) one 
obtains a deformed equation describing the $p_1$-threshold: 
\begin{equation} 
p_{1,th} \simeq {(m_2 + m_3)^2 - m_1^2 \over 4 \epsilon} 
+ \eta {p_{1,th}^{2+\alpha} \over 4 \epsilon E_p^\alpha} \left(  
{m_2^{1+\alpha} + m_3^{1+\alpha} \over (m_2 + m_3)^{1+\alpha}} -1 \right) 
~. 
\label{lithresh2} 
\end{equation} 
where we have included only the leading corrections (terms suppressed 
by both the smallness of $E_{p}^{-1}$ and the smallness of $\epsilon$ 
or $m$ were neglected). 
 
\subsection{Phenomenology} 
Early phenomenological interest in the proposal (\ref{dispone}) came 
from studies based on time-of-flight 
analyses~\cite{grbgac,schaef,billetal} of photons associated with 
gamma-ray bursts or with Mk 421.  According to (\ref{dispone}) (and 
assuming that there is no leading-order deformation of the standard 
relation $v = dE/dp$) one would predict~\cite{grbgac,aemn1} 
energy-dependent relative delays between the times of arrival of 
simultaneously emitted massless particles: 
\begin{equation} 
{\Delta T \over T} =  
\eta {(\alpha + 1) \over 2}{E'^\alpha - E^\alpha \over E_p^\alpha} 
~, 
\label{velox} 
\end{equation} 
where $T$ is the (average) overall time of travel of simultaneously 
emitted massless particles and $\Delta T$ is the relative delay 
between the times of arrival of two massless particles of energies $E$ 
and $E'$.  The fact that such time delays have not yet been observed 
allows us to set bounds on the $\alpha,\eta$ parameter space.  In 
particular, data showing (approximate) simultaneity of arrival of 
TeV photons from Mk 421 were used~\cite{billetal} to set the 
bound $|\eta| < 3 {\cdot} 10^2$ for $\alpha=1$.  The same data were used 
in Ref.~\cite{polonpap} to set a more general $\alpha$-dependent bound 
on $\eta$. 
 
We combine these existing bounds with the assumption that indeed the 
UHECR and TeV-$\gamma$ threshold anomalies are due to LID 
(\ref{dispone}).  The fact that the scale $E_{p}$ is very high might 
give the erroneous impression that the new term going like 
$p_{1,th}^{2+\alpha} /E_{p}^\alpha$ present in Eq.~(\ref{lithresh2}) 
could always be safely neglected, but this is not the 
case~\cite{ita,sato,kifu,kluz,Protheroe_Meyer,AmePiran00}.  For given 
values of $\alpha,\eta$ one finds values of $\epsilon$ that are low 
enough for the ``threshold anomaly''~\cite{AmePiran00}
(displacement of the threshold) 
to be significant.  For certain combinations $\alpha,\eta,\epsilon$ 
the threshold completely disappears, {\it i.e.} Eq.~(\ref{lithresh2}) 
has no solutions.  Assuming (\ref{lithresh2}) one would predict 
dramatic departures from the ordinary expectations of Lorentz 
invariance; in particular, if $\alpha \sim - \eta \sim 1$, according 
to (\ref{lithresh2}) one would expect that the Universe be transparent 
to TeV photons.  The corresponding result obtainable in the UHECRs 
context would imply that the GZK cutoff could be violated~\cite{kifu} 
even for much smaller negative values of $\eta$.  Positive values of 
$\eta$ would shift the thresholds in the opposite direction 
({\it e.g.} they would imply an even stricter limit than the GZK one) 
and are therefore not consistent with the hypothesis that UHECR 
and TeV-$\gamma$ threshold anomalies be due to LID (\ref{dispone}). 
 
In Figure~1 we provide a quantitative description of the region of the 
$\alpha,\eta$ parameter space which would provide a solution to both 
the UHECR and TeV-$\gamma$ threshold anomalies while satisfying the 
time-of-flight constraints~\cite{schaef,billetal} that are still 
consistent with ordinary Lorentz invariance.  The curve describing the 
time-of-flight constraints was obtained using the information that 
there is~\cite{billetal} an upper bound of order 200 seconds to the 
difference in time of arrivals of 2TeV photons and 1TeV photons 
simultaneously emitted by Mk 421 (at redshift of 0.031).  The two 
threshold-anomaly curves reported in Figure~1 were obtained using 
Eq.~(\ref{lithresh2}) with $m_1=0$ and $m_2=m_3 = 5 {\cdot} 10^5$eV 
(TeV photons $\gamma + \gamma \rightarrow e^+ + e^-$ threshold 
analysis) and with $m_1 = m_2 = 9.4 {\cdot} 10^8$eV and $m_3= 1.4 
{\cdot} 10^8$eV (UHECR $p + \gamma \rightarrow p + \pi$ threshold 
analysis\footnote{The dominant contribution to the GZK cutoff actually 
  comes from the $\Delta$ resonance, so one might find appropriate to 
  replace the sum of the proton mass and the pion mass with the mass 
  of the $\Delta$ in the UHECR threshold formula.  However, the 
  difference between $m_\Delta$ and $m_p + m_\pi$ would only introduce 
  a relatively small correction in our UHECR limit which is not our 
  dominant lower limit (a much stricter limit comes from the 
  TeV-$\gamma$ anomaly). Moreover, once the contribution to GZK from 
  the $\Delta$ is avoided one would still have a (weakened) GZK cutoff 
  from non-resonant photopion production and this would anyway lead to 
  the limit we use.}).  In light of the analysis of the experimental 
situation provided in Sections~2 and 3, we obtained the UHECRs curve 
by requiring sufficient LID to explain the factor-6 threshold shift $5 
{\cdot} 10^{19}$eV$ \rightarrow 3 {\cdot} 10^{20}$eV, while for the 
TeV photons curve we required a factor-2 threshold shift $10$TeV 
$\rightarrow 20$TeV.  Even though the shift is more significant in the 
UHECR context, it is the requirement to explain the TeV-$\gamma$ 
threshold anomaly that provides a more stringent constraint, as one 
should expect since in our LID, which is motivated by Planck-scale 
physics, the violation of ordinary Lorentz invariance is suppressed by 
some power of the ratio $E/E_p$.

\begin{figure}[p] 
\begin{center} 
\epsfig{file=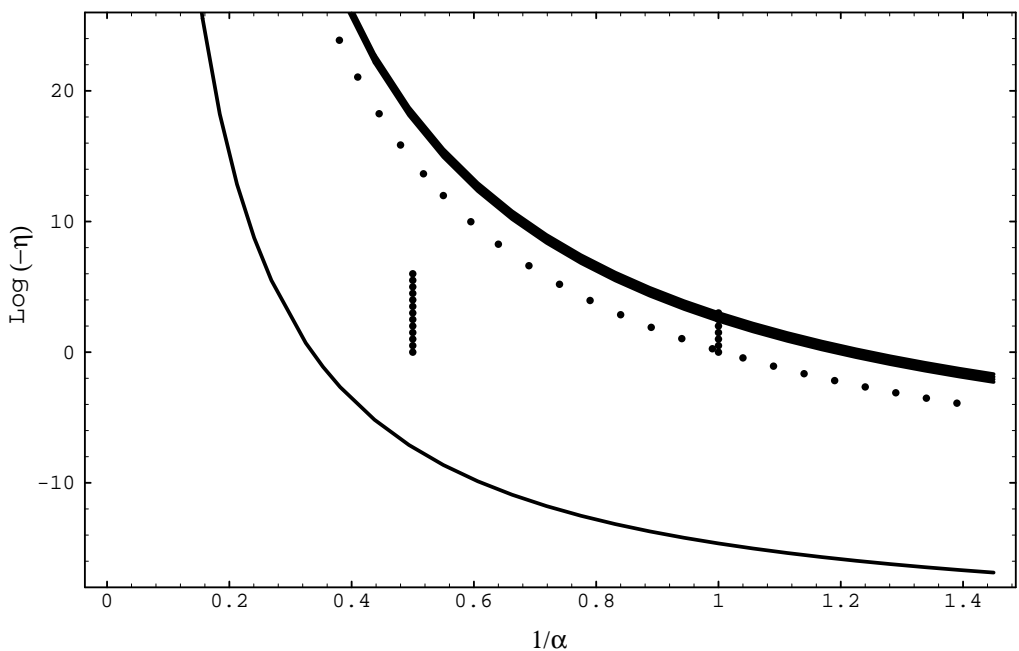,height=14truecm}\quad 
\end{center} 
\caption{The region of the $\alpha,\eta$ parameter space that provides a
solution to both the UHECR and TeV-$\gamma$ threshold anomalies while
satisfying the time-of-flight upper bound on LID.  Only negative
values of $\eta$ are considered since this is necessary in order to
have upward shifts of the threshold energies, as required by the
present paradoxes.  The solid thick line describes the time-of-flight
upper bound.  The region above this line is excluded.  The solid thin
line and the dotted line describe the lower bound on LID obtained from
the present UHECR (solid thin line) and TeV-$\gamma$ (dotted line)
threshold anomalies.  The anomalies disappear in the region above the
lines. Within the narrow region between the dotted line and the solid
thick line the time of flight constraint is satisfied and both
anomalies are resolved.  The two vertical segments at $\alpha = 1$ and
at $\alpha = 2$ ({\it i.e.} at $1/\alpha = 1/2$) correspond to the two
favored quantum-gravity scenarios.  The behaviour of the curves for
upper and lower bounds on LID with respect to the bottom-left corner
of the frame can be understood by noticing that at a fixed $\alpha$
ordinary Lorentz invariance can be reached taking 
the $\eta \rightarrow 0$ limit, while at fixed $\eta$ this requires
taking the $\alpha \rightarrow \infty$ 
({\it i.e.} the $1/\alpha \rightarrow 0$) limit.}
\label{fig1} 
\end{figure}

Considering the diverse origin and nature of the three relevant 
classes of experimental data that we are considering,  
the fact that there is a 
region of the $\alpha,\eta$ parameter space consistent with all these 
constraints is non-trivial, and this in turn provides encouragement for 
the interpretation of the threshold anomalies as manifestations of LID. 
Moreover, it is quite striking that this 
region of parameter space, in spite of being relatively small, does 
contain one of the two mentioned quantum-gravity-motivated scenarios: 
$\alpha =1$ and $1 < \eta < 10^{3}$. The other 
quantum-gravity-motivated scenario, the one  
with $\alpha =2$ and $1 < \eta < 10^{6}$,  
is outside the relevant region of parameter space, 
being consistent with the absence of relative time delays and  
the UHECR threshold anomaly but 
being inconsistent with threshold anomaly for multi-TeV photons. 
 
Concerning the consistency of the interpretation of the threshold 
anomalies as manifestations of LID it is also important to observe 
that the modified dispersion relation (\ref{dispone}), in spite of 
affecting so significantly the GZK and TeV-$\gamma$ thresholds, does not 
affect 
significantly the processes used for the detection of the relevant 
high-energy particles.  For the significance of the threshold 
modification a key role is played, as evident from equation 
(\ref{lithresh2}), by the smallness of the energy of the background 
photons.  The effect of (\ref{dispone}) on atmospheric interactions of 
the relevant high-energy particles is instead suppressed by the fact 
that in these atmospheric interactions the ``targets", nuclei or electrons, 
have energies much higher than those of the background photons.           

\vfil  
\eject 
 
\newpage  

\section{A more general LID formalism} 
 
Having shown that the simple two-parameter family of 
Lorentz-invariance-violating dispersion relations (\ref{dispone}) 
provides a solution of the UHECRs and TeV-$\gamma$ threshold 
paradoxes, we turn now to a more general five-parameter LID 
formulation.  The motivation for this formulation comes primarily from 
theory work on short-distance (so called, ``quantum gravity") 
properties of space-time, in which modifications of space-time 
symmetries are encountered quite naturally.  In particular, 
quantum-gravity effects inducing some level of nonlocality or 
noncommutativity would affect even the most basic flat-space 
continuous symmetries, such as Lorentz invariance.  
This has been recently emphasized in various quantum-gravity 
approaches~[31-33,39-50]
based on critical or noncritical string theories, noncommutative 
geometry or canonical quantum gravity.  While we must be open to 
the possibility that some symmetries are completely lost, 
it appears plausible that some of them are not really lost
but rather replaced by a Planck-scale-deformed version. 
Some mathematical frameworks which could consistently describe such
deformations have emerged in the mathematical-physics 
literature~\cite{gackpoinplb,kpoinap,gacmaj,lukipap,majrue}.  An 
example of these structures is discussed in Appendix~A. 
 
\subsection{The five-parameter formalism} 
 
The fact that a simple two-parameter family of 
Lorentz-invariance-violating dispersion relations (\ref{dispone}) is 
consistent with all available data is of course of encouragement for 
the LID hypothesis, but, especially since relevant data are expected 
to improve rapidly in the coming years, it is also important to 
establish how much room for generalizations of (\ref{dispone}) is 
available in the general framework of Planck-scale-induced LID. 
 
One way to generalize (\ref{dispone}) would involve attributing 
different independent values of $\alpha,\eta$ to different particles. 
We shall not pursue this (however phenomenologically viable) 
possibility, since the focus of the present article is on deformations 
of Lorentz symmetry which could be induced by non-trivial 
space-time structure, and such deformations would 
most likely treat ``democratically" all particles.  In any case, it is 
clear that models attributing different independent values of $\alpha$ 
and $\eta$ to each particle end up having a very large number of free 
parameters and available data will not be very effective in 
constraining such models.  We shall come back to this point in 
Section~6, where we consider the alternative (Planck-scale 
independent) Coleman and Glashow~\cite{colgla} scheme for 
Lorentz-invariance-violation.  In fact, that scheme corresponds to the 
choice $\alpha = 0$ and an independent value of $\eta$ for each 
particle. 
 
Another way to generalize the dispersion relation (\ref{dispone}) is to 
include other deformation terms.  
In a space-time with some non-trivial structure  
at distances of order $E_p^{-1}$ one could  
expect that probes with energy much smaller than $E_p$ 
should obey a dispersion relation of type: 
\begin{equation} 
E^2 - p^2 - m^2 = F(E,p,m;E_p) 
~, 
\label{toogeneral} 
\end{equation} 
where $F$ is some general function with units of mass (or energy) 
squared and such that $F \rightarrow 0$ for $E_p \rightarrow \infty$. 
Actually, in studies, such as ours, looking only for the leading 
correction, one of the arguments of $F$ can be suppressed: one makes a 
subleading error by using $E^2 - p^2 - m^2 = 0$ to express one of the 
variables in $F$ in terms of the other variables.  One could for 
example express $F$ as a function of $p$ and $m$ only: $F(p,m;E_p)$. 
Moreover, the fact that we are only looking for the leading correction 
in the high-energy regime\footnote{It is perhaps worth emphasizing 
  that the low-energy expansion of $F(p,m;E_p)$ may look quite 
  different from its corresponding high-energy expansion.  In the 
  high-energy regime ($p \gg m$) the premium is on the leading 
  dependence on $p$ while in the low-energy regime ($p \ll m$) the 
  leading dependence on $m$ is dominant.}  allows us to approximate 
$F$ with its leading (if any) power dependence on $E_p$ and (within a 
given power dependence on $E_p$) leading dependence 
on $p$: $F(p,m;E_p) \simeq \eta \, p^{2
+\alpha-\sigma} \, m^{\sigma} \, E_p^{-\nu}$.  
In the high-energy regime there is therefore scope for 
considering the three-parameter family of dispersion 
relations
\begin{equation} 
E^2 - p^2 - m^2 \simeq \eta {\cdot} p^{2+\alpha-\sigma} {\cdot} 
m^{\sigma} {\cdot} E_p^{-\alpha}  
~, 
\label{dispfull} 
\end{equation} 
where, of course, it is understood 
that $m^{\sigma}=1$ whenever $\sigma=0$, even when $m=0$.
(The parameter $\sigma$ has been introduced to characterize the
type of dependence of the deformation term on the mass $m$,
and therefore in our notation there is the implicit prescription
that $m^{\sigma} \rightarrow 1$ in the formally ambiguous 
combined limit $\sigma \rightarrow 0$, $m \rightarrow 0$.)

Besides the structure of the dispersion relation 
a LID can also affect the law of sum of momenta. 
Since our emphasis is here on the phenomenology of LIDs,  
rather than on their formal/mathematical analysis, we limit our discussion 
of the motivation for this type of effect to the example  
of non-commutative geometry (the ``$\kappa$-Minkowski space-time") 
considered in Appendix~A. 
As that example clarifies,  
it is natural to consider a two-parameter 
class of modifications of the law of sum of (parallel) momenta 
$K_1 + K_2 \rightarrow K_1 + K_2  
+ \delta (K_1 K_2)^{(1+\beta)/2} E_{p}^{-\beta}$. 
For our threshold analyses this corresponds to 
\begin{equation} 
p_1 - \epsilon \rightarrow  
p_1 - \epsilon - \delta {(p_1 \epsilon)^{(1+\beta)/2} \over E_{p}^\beta} 
\simeq p_1 - \epsilon  
~,~~~ 
p_2 + p_3 \rightarrow  
p_2 + p_3 + \delta {(p_2 p_3)^{(1+\beta)/2} \over E_{p}^\beta} 
~. 
\label{psumrule} 
\end{equation} 
 
Overall we consider a five-parameter space: $\alpha,\eta,\sigma$  
for the dispersion relation and $\beta,\delta$ for the description  
of possible deformations (\ref{psumrule}) of 
the law of addition of momenta.   
The analysis reported in the preceding Section corresponds of course 
to the $\sigma \rightarrow 0$, $\delta \rightarrow 0$ limit  
of this more general five-parameter 
($\alpha,\eta,\sigma,\beta,\delta$) phenomenology. 
 
As appropriate for the present preliminary status of the experimental 
situation and the fact that the two-parameter phenomenology analyzed 
in the previous Section turned out to give a fully satisfactory 
description of the data, we shall only provide here a preliminary and 
partial exploration of the enlarged five-parameter space.  Our 
exploration of this parameter space will also be more detailed in some 
directions and less detailed in others.  In particular, we shall limit 
our analysis to two classes of scenarios, one with $\sigma = 0$ and 
one with $\sigma=2,\delta=0$.  This will be sufficient for a 
qualitative understanding of how different portions of our 
five-parameter space compare with the present experimental situation. 

Retaining the leading corrections in $E_{p}^{-1}$, the threshold 
analysis in the general five-parameter 
($\alpha,\eta,\sigma,\beta,\delta$) LID scenario leads to the 
threshold equation:\footnote{Note that actually the threshold is not 
necessarily anomalous; in particular, as we already observed in 
Ref.~\cite{AmePiran00}, when $\alpha=\beta=1$, $\sigma=0$
and $\eta=-\delta$ there is a 
cancellation and the deformed symmetries lead to the same threshold 
equation obtained with undeformed symmetries.} 
\begin{eqnarray} 
p_{1,th} \!\!\!&\simeq&\!\!\! {(m_2 + m_3)^2 - m_1^2 \over 4 \epsilon} 
+ \eta {p_{1,th}^{2-\sigma} \over 4 \epsilon}  
\left( {m_2^{1+\alpha} + m_3^{1+\alpha} \over (m_2 + m_3)^{1+\alpha-\sigma}} 
- m_1^\sigma \right) \left({p_{1,th}\over E_p}\right)^\alpha  
\label{newmast} \\ 
\nonumber 
& &- \delta {p_{1,th}^{2} \over  
2 \epsilon}  
\left( {\sqrt{m_2 m_3} \over m_2 + m_3} \right)^{1+\beta} 
 \left({p_{1,th} \over E_p}\right)^\beta ~. 
\end{eqnarray} 
In the following sections we apply this equation to several specific cases.  
To simplify the discussion we provide here explicit expressions for 
the threshold for photopion production: 

\eject

\begin{eqnarray} 
E_{GZK,th}  \!\!\!& \simeq &\!\!\! {7 {\times} 10^{19}~ {\rm eV} \over  
\epsilon/ 0.001 {\rm eV}  } 
\left[ 1  +  \eta ~ 
10^{22.2-10.9\sigma-8.15 \alpha}  
\left( (0.87^{1+\alpha}+ 0.13^{1+\alpha}) 1.15^\sigma  -1 \right)~~~~~~
 \right. \nonumber \\ & &~~~~~~~~~~~~~~~~~~~~~~~~
~~\left({E_{GZK,th}\over 7 {\times} 10^{19} {\rm eV}}\right)^{2-\sigma} 
\left({E_{GZK,th}/7 {\times} 10^{19} {\rm eV} 
\over E_p/10^{19} {\rm GeV} }\right)^\alpha 
\label{neweq1} \\ \nonumber 
& &~~~~~~~~~~~~~~~~~~ - ~\delta~ \left. 10^{22.1-8.15\beta}  
\left({E_{GZK,th}\over 7 {\times} 10^{19}{\rm eV} }\right)^2  
\left({E_{GZK,th}/7 {\times} 10^{19} {\rm eV}\over E_p/10^{19} 
{\rm GeV}}\right)^\beta  
\right]
 ~,  
\end{eqnarray} 
and for pair creation threshold: 
\begin{eqnarray} 
E_{\gamma,th} \!\!\!& \simeq &\!\!\! 
{25 ~ {\rm TeV} \over \epsilon/ 0.01 {\rm eV}  } 
\left[ 1 + \eta ~    10^{14.8 -7.7 \sigma - 14.6 \alpha}
\left(2^{\sigma-\alpha} - \delta^K_{\sigma,0}\right) 
\left({E_{\gamma,th}\over 25 {\rm TeV}}\right)^{2-\sigma} 
\left({E_{\gamma,th}/25 ~ {\rm TeV}\over E_p/10^{19} 
{\rm GeV}}\right)^\alpha \right. \label{neweq2}  \nonumber \\ 
& &~~~~~~~~~~~~~~~ - \left. \delta ~  10^{14.8 - 14.9 \beta} 
\left({E_{\gamma,th}\over 25 {\rm TeV}}\right)^{2} 
\left({E_{\gamma,th}/25 ~ {\rm TeV}\over E_p/10^{19} 
{\rm GeV}}\right)^\beta  
\right] ~,
\end{eqnarray} 
where we found convenient to introduce 
the ``Kronecker delta", here denoted with $\delta^K$
to differentiate it from our parameter $\delta$,
to compactly write this equation consistently
with our conventions for the $m_1 \rightarrow 0$ 
limit. [In deriving Eq.~(\ref{neweq2}) from
Eq.~(\ref{newmast}) it is necessary to take into account
that, consistently with the conventions and notations we 
introduced (see, in particular, the comments made 
immediately after Eq.~(\ref{dispfull})),
in the limit $m_1 \rightarrow 0$ the term $m_1^\sigma$ must be handled
according to $m_1^\sigma \rightarrow 0$ if  $\sigma \ne 0$
and according to $m_1^\sigma \rightarrow 1$ if  $\sigma = 0$.
Of course, the reader can verify by direct calculation that this 
prescription gives the correct threshold conditions that follow
from Eq.~(\ref{dispfull}) in the two cases $\sigma = 0$ 
and $\sigma \ne 0$, and reproduces the threshold 
condition (\ref{lithresh2}) obtained in the preceding Section
(which was devoted to the case $\sigma = 0, \delta=0$).]

\subsection{Phenomenology with $\sigma=2,\delta=0$} 
 
In the case $\sigma=2,\delta=0$ there is no deformed law of addition 
of momenta and the threshold equation takes the form  
\begin{equation} 
p_{1,th} \simeq {(m_2 + m_3)^2 - m_1^2 \over 4 \epsilon} 
+ \eta {p_{1,th}^{\alpha} \over 4 \epsilon E_p^{\alpha}}  
\left({m_2^{1+\alpha} + m_3^{1+\alpha} \over (m_2 + m_3)^{\alpha-1}}  
- m_1^2 \right) 
~. 
\label{newmastwithm} 
\end{equation} 
 
For $\sigma \ne 0$ the LID term in (\ref{dispfull}) vanishes for 
massless particles.  Thus, in general in all $\sigma \ne 0$ cases 
(like the one we discuss in this Subsection) the time of flight 
constraints~\cite{schaef,billetal} do not limit the LID parameters. 
 
The constraints obtainable by interpreting the UHECR and TeV-$\gamma$ 
threshold anomalies as manifestations of LID suggest that this 
interpretation is quite unnatural in the case $\sigma=2,\delta=0$. 
The condition that both threshold be pushed upwards leads to the 
constraints $\eta >0$, $\alpha < 1.195$.  Moreover, in order to 
describe the threshold anomaly for multi-TeV photons one should also 
make the awkward requirement $\eta > 10^{15 \alpha}$.   
Having provided in the previous section 
an elegant solution of the threshold paradoxes using $\sigma=0$ we do 
not pursue further this scenario which appears to require a higher level 
of fine tuning. 
 
Scenarios with $\sigma \ne 0$ might regain some interest if there are 
significant new developments in the understanding of the threshold 
anomalies that will point in this direction. In the present 
experimental and theoretical situation we find appropriate to make in 
the following the assumption that $\sigma=0$. 
 
\subsection{General aspects of the phenomenology with $\sigma=0$} 
 
For $\sigma=0$ one is left with a four-parameter space on which 
significant information can be gained by combining data on possible 
time of flight delays, which will only constrain, through the 
prediction (\ref{velox}), the parameters $\alpha,\eta$, and data on 
the threshold anomalies, which, through (\ref{newmast}), are relevant 
for all four parameters $\alpha,\eta,\beta,\delta$. 
 
It is important to observe that positive (discovery) results on both 
the thresholds and the time delays would allow to determine the values 
of all four parameters.  If eventually the mentioned time delays are 
actually observed, and if they are observed in signals from a 
collection of sources diverse enough to allow the determination of the 
energy dependence of the time delays, we would then be able to use 
(\ref{velox}) to fix $\alpha$ and $\eta$.  Then, knowing $\alpha$ and 
$\eta$, a determination of the thresholds could be used to fix $\beta$ 
and $\delta$. 
 
While waiting for these eventual discoveries, one can use the present 
upper limits on LID in relative time delays and (preliminary evidence 
of) lower limits on LID in threshold anomalies to reduce the allowed 
portion of the four-parameter space.  We subdivide the discussion of 
this type of phenomenological analysis in three 
cases: $\alpha < \beta$, $\alpha = \beta$ and $\alpha > \beta$. 
 
\subsection{Phenomenology with  $\alpha < \beta$ (and $\sigma=0$)} 
 
The case $\alpha < \beta$ (and $\sigma=0$) is essentially analogous to 
the case considered in the preceding Section with the two-parameter 
$\alpha,\eta$ phenomenology.  In fact, for $\alpha < \beta$ the 
threshold corrections associated with the deformation of the law of 
addition of momenta are suppressed by factors of 
order $(E/E_p)^{\beta  - \alpha}$ with respect to 
the threshold corrections associated with 
the deformed dispersion relation.  The constraints derived for 
$\alpha,\eta$ in the preceding Section would still be valid and, as 
long as we have only lower or upper limits (rather than definite 
discoveries), no constraint could be put on $\beta,\delta$. 
 
\subsection{Phenomenology with $\alpha = \beta$ (and $\sigma=0$)} 
 
For $\alpha = \beta$ (and $\sigma=0$) the upper limit on 
time-of-flight LID still constrains only $\alpha,\eta$, but the 
constraints on $\alpha,\eta$ obtainable by interpreting the UHECR and 
TeV-$\gamma$ threshold anomalies as manifestations of LID are weakened 
by allowing also a deformed law of addition of momenta. In practice 
the parameters $\alpha,\eta$ and $\beta,\delta$ can in a sense ``share 
the burden" of explaining the threshold anomalies.  To illustrate this 
mechanism we show in Figure~2 the constraints on $\eta,\delta$ that 
are obtained for $\alpha = \beta = 1$ (and $\sigma=0$). 
 
\subsection{Phenomenology with $\alpha > \beta$ (and $\sigma=0$)} 
 
For $\alpha > \beta$ (and $\sigma=0$) the threshold corrections 
associated with the deformed dispersion relation are suppressed by 
factors of order $(E/E_p)^{\alpha - \beta}$ with respect to the 
threshold corrections associated with the deformation of the law of 
addition of momenta.  Therefore the interpretation of the UHECR and 
TeV-$\gamma$ threshold anomalies as manifestations of LID imposes 
constraints (lower bounds on LID) on the parameters $\beta,\delta$. 
As always, the upper limit on time-of-flight LID constrains only 
$\alpha,\eta$.  It is worth noticing that if future data should 
indicate that there is no LID relative time-delay effect but there are 
LID threshold anomalies this scenario with $\alpha > \beta$ would 
become favored. 
 
 
\begin{figure}[p] 
\begin{center} 
\epsfig{file=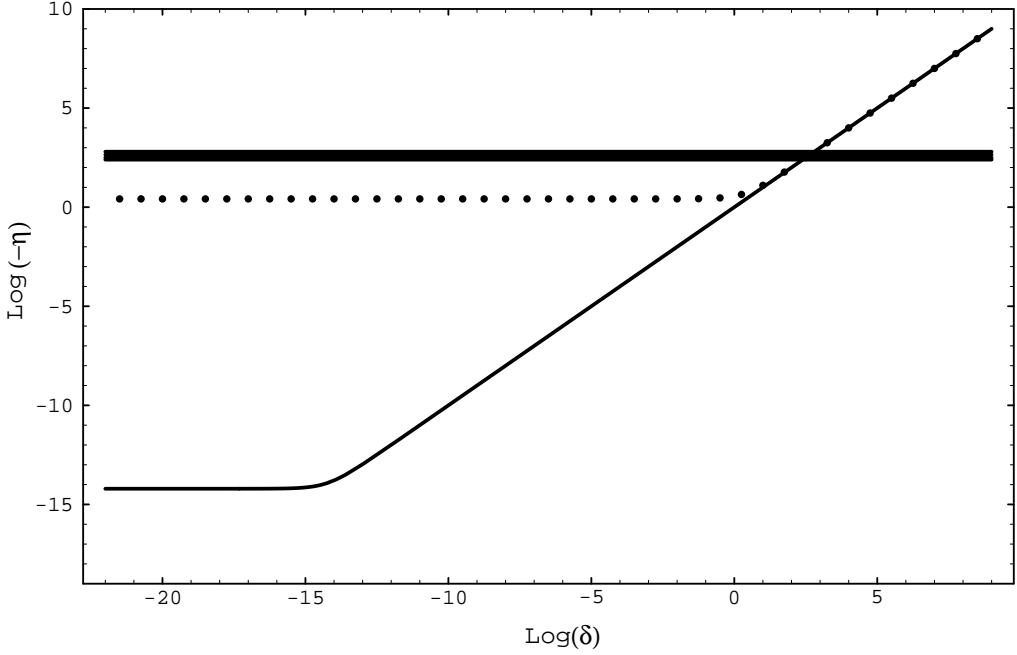,height=14truecm}\quad 
\end{center} 
\caption{A two-dimensional slice through the five-dimensional parameter
space. Shown is the $\eta <0$ and $\delta >0$ region 
for $\alpha = \beta =1$ and $\sigma =0$.  
Analogous considerations (with exchange of roles between $\eta$ 
and $\delta$) also apply to the corresponding $\eta >0$ 
and $\delta <0$ region.  
As in Figure~1, the thick solid line describes the
time-of-flight upper bound, while the tentative lower bounds on LID 
that can be obtained from the present UHECR and TeV-$\gamma$ threshold anomalies
are described by the thin solid line and the dotted line respectively.  
Notice that for $\delta < 10^{-14}$ the range of
allowed $\eta$ values is almost unaffected by $\delta$, while values
of $\delta$ such that $\delta > 10^{2.4}$ are not consistent with the
working hypothesis of the present Article: the tentative
threshold-anomaly lower bound is higher than the time-of-flight upper
bound for $\delta > 10^{2.4}$.}
\label{fig2} 
\end{figure} 
 
 
 
\begin{figure}[p] 
\begin{center} 
\epsfig{file=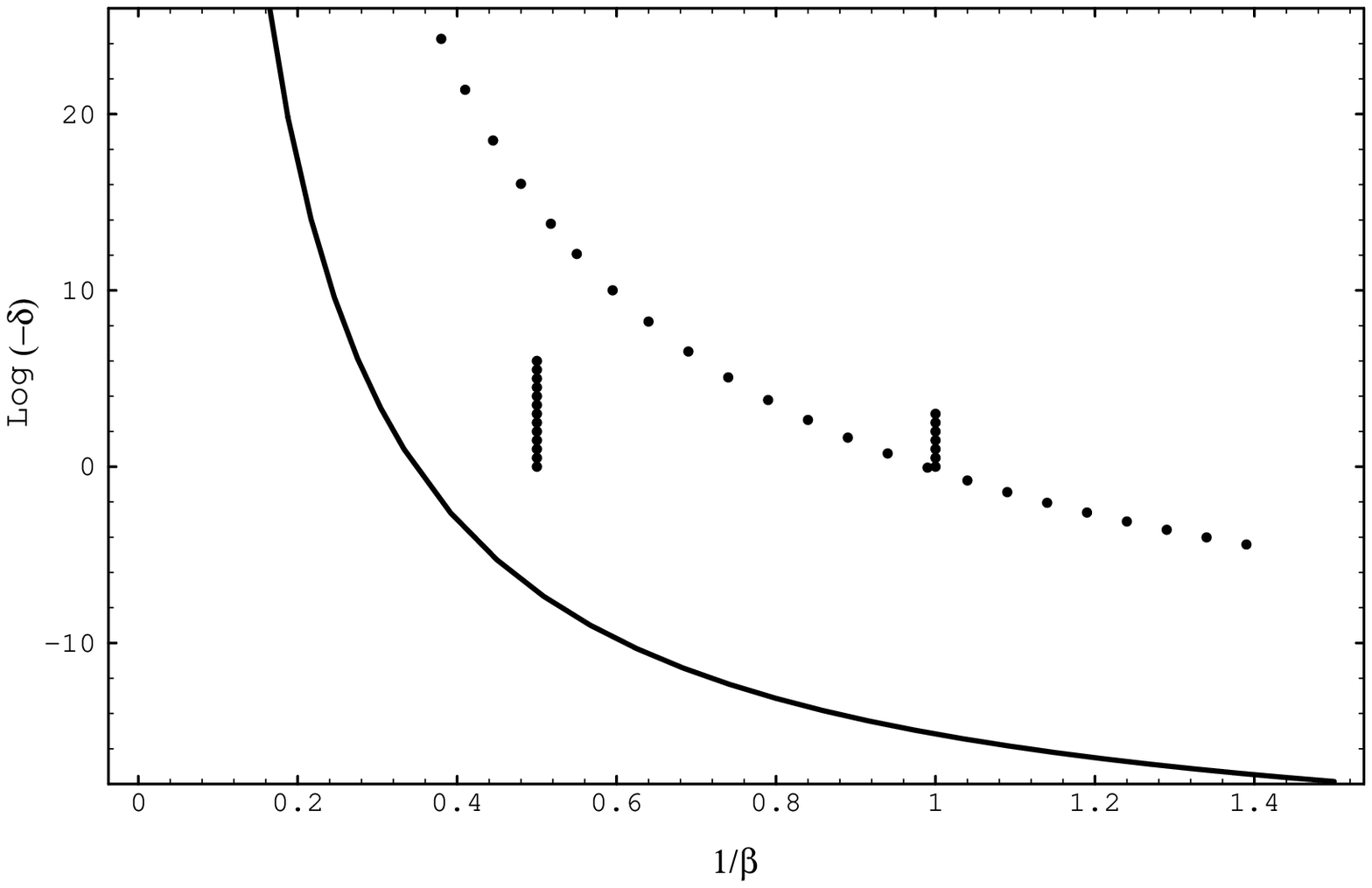,height=14truecm}\quad 
\end{center} 
\bigskip 
\caption{The region of the $\beta,\delta$ parameter space 
that provides a solution to both the UHECR and TeV-$\gamma$ threshold
anomalies for $\alpha > \beta$.  
Only negative values of $\delta$ are considered since,
when $\alpha > \beta$, this is necessary in order to have upward
shifts of the threshold energies, as required by the present paradoxes.
As in Figure~1, the tentative lower bounds on LID that can be obtained 
from the present UHECR and TeV-$\gamma$ threshold anomalies
are described by the thin solid line and the dotted line respectively.}
\label{fig3} 
\end{figure} 

Figure~3 depicts the limits on $\beta,\delta$ that follow, when 
$\alpha > \beta$, from interpreting the UHECR and TeV-$\gamma$ 
threshold anomalies as manifestations of LID.  The limits on 
$\alpha,\eta$ due to the upper limit on time-of-flight LID are still 
the same as in Figure~1 (but, as just mentioned, the two 
threshold-anomaly curves in Figure~1 do not apply when $\alpha > 
\beta$). 

\vfil  
\eject 
 
\section{Comparison with the Coleman-Glashow scheme} 
 
Coleman and Glashow~\cite{colgla} have recently introduced a different 
scheme (denoted CG scheme hereafter) for violation of Lorentz 
invariance.  Modifying the elementary particles Lagrangian they 
suggest a scheme in which there is a different maximum attainable 
velocity, $c_a$, for each particle.  The relevant dispersion relations 
take the form 
\begin{equation} 
E^2  - p^2  c_a^2 =   m^2 c_a^4 ~, 
\label{colglaeq} 
\end{equation} 
where the index $a$ labels the particle.  In the language developed in 
Sections~4 and 5 these dispersion relations (\ref{colglaeq}) involve 
two terms, one with $\alpha = 0$ and $\sigma=0$ and the other with 
$\alpha = 0$ and $\sigma=2$. The particle-dependence of $c_a$ 
could be described by allowing for a different independent value  
of $\eta$ for each fundamental particle.   
At high energies, in which we are interested, the $\alpha = 0,\sigma=0$  
term dominates and $\eta_a = c^2 - c_a^2 \approx 2 c (c-c_a)$.   
The condition $\alpha = 0$ reflects the fact 
that the CG scheme is not motivated by Planck-scale physics.  The 
possibility for each particle to get its own independent value of 
$\eta$ reflects the fact that this scheme is not intended as a 
description of deformations of Lorentz invariance due to non-trivial 
short-distance space-time structure. (If a deformation of Lorentz 
symmetry is induced by the structure of space-time we expect that it 
would affect all particles in the same way. Such a symmetry 
deformation might allow for a dependence of the correction terms on 
the mass and the spin of the particle but the parameters of the model 
should not depend on the mass, spin or other quantum numbers of the 
particles.)

Using again the language we developed in Sections~4 and 5 one can also 
give an intuitive characterization of the way in which the CG scheme 
and the scheme considered here are alternative to one another as 
strategies for obtaining threshold anomalies.  In fact, in that 
language one could describe undeformed thresholds\footnote{The fact 
  that there are no UHECR and TeV-$\gamma$ threshold anomalies in our 
  scheme for $\alpha = 0,\sigma=0,\delta=0$ can be easily derived 
  directly from the corresponding dispersion relation. This is also 
  implicit in Figure~1, which shows that $|\eta| \rightarrow \infty$ as 
  $\alpha \rightarrow 0$. Also notice that undeformed thresholds are 
  not only obtained for $\alpha = 0,\sigma=0,\delta=0$:  
  the thresholds become undeformed also, for example,  
  in the limit $\alpha \rightarrow \infty$ approached  
  keeping $\sigma=0,\delta=0$.  However the 
  undeformed-threshold point $\alpha = 0,\sigma=0,\delta=0$ is best 
  suited for a comparison between our scheme and the CG scheme.}  as 
associated with $\alpha = 0,\sigma=0,\delta=0$, independently of the 
value of $\eta$.  This corresponds to the 
fact that in the CG scheme there are of course no threshold anomalies 
if all $c_a$'s take the same value ($c_a = c - \eta/(2 c)$). Threshold 
anomalies are generated in the CG scheme by deforming the 
threshold conditions in the direction that corresponds to 
keeping $\alpha = 0,\sigma=0,\delta=0$ but allowing different 
independent values of $c_a$ for each fundamental particle.  On the 
contrary, in our scheme the threshold anomalies are 
obtained by allowing for deviations from $\alpha = 0,\sigma=0,\delta=0$ 
while keeping a single (particle-independent) $\eta$. 
 
In light of these comments it is not surprising that threshold 
anomalies within the CG scheme take the characteristic ``$c_a - c_b$" 
dependence.  In particular, as already observed in Ref.~\cite{colgla}, 
the description of the UHECR threshold anomaly requires (together with 
conditions on $c_\Delta - c_p$) that $c_\pi - c_p > 10^{-24}$. 
($c_\pi$ and $c_p$ are the $c_a$'s for pions and protons 
respectively.)  We observe that a resolution of the TeV-$\gamma$ 
threshold anomaly within the CG scheme requires the additional 
condition $c_e - c_\gamma > 5 {\cdot} 10^{-16}$.  This combines with the 
absence~\cite{colgla} of vacuum Cerenkov radiation by electrons with 
energies up to 500GeV in such a way that $c_e - c_\gamma$ is bound to 
$5 {\cdot} 10^{-16} < c_e - c_\gamma < 5 {\cdot} 10^{-13}$.  There is therefore 
a relatively narrow range of allowed values for $c_e - c_\gamma$ just 
like\footnote{Note however that, while the TeV-$\gamma$ threshold 
  anomaly is used in both, not all the experimental constraints used 
  by the two phenomenological analysis are the same.  In particular, 
  the time-of-flight upper bound on LID was not used to establish 
  $5 {\cdot} 10^{-16} < c_e - c_\gamma < 5 {\cdot} 10^{-13}$.}  we found in 
Section~4 a relatively narrow allowed region of the $\alpha,\eta$ 
parameter space. 
 
One important difference between the two schemes is that in our 
Planck-scale-motivated LID the allowed region of parameter space is 
found exactly where quantum-gravity intuition would have sent us 
searching for new physics, while in the CG scheme values 
of $c_e - c_\gamma$ in the 
range $5 {\cdot} 10^{-16} < c_e - c_\gamma < 5 {\cdot} 10^{-13}$ 
do not have any special significance.  Another important 
difference between the two schemes is that while the same 
$\alpha,\eta$ parameters of our scheme for LID are also constrained by 
UHECR threshold data, in the CG scheme $c_e - c_\gamma$ does not play 
any role in the equation for the UHECR threshold and vice versa.  Any 
future development in the UHECR threshold data would leave $c_e - 
c_\gamma$ unaffected.  On the contrary, the plausibility of the 
Planck-scale-motivated LID will be strongly affected by future UHECR 
threshold data: if the lower limit on the threshold continues to be 
pushed higher the overall consistency and appeal of the LID model 
would increase, while the discovery of the threshold not much higher 
than the present $3 {\cdot} 10^{20}$eV lower limit would (unless the 
TeV-$\gamma$ threshold anomaly is eventually understood as a result of 
systematic errors) rule out the model considered in Section~4. 
 
While the scheme considered here is more tightly constrained by 
high-energy data (because all high-energy data set constraints to the 
same few space-time related parameters), the CG scheme is constrained 
more tightly than ours by low-energy data.  The parameters we 
considered in the present Article, dealing exclusively with the 
high-energy regime, are practically unconstrained by low-energy data 
since, as discussed in Section~5, the LID we considered here might 
emerge in quantum gravity as the leading order in the high-energy 
expansion of an analytic function whose low-energy expansion looks 
quite different.  On the contrary the CG scheme takes a fixed 
(energy-independent) value of its parameters $c_a$ and therefore 
high-energy and low-energy data can be combined to obtain stricter 
limits. 
 
\section{Summary and outlook} 
 
In the present Article we took as working assumption that the UHECR
and TeV-$\gamma$ threshold anomalies do not have a simple explanation
(whereas, especially for the case of TeV photons, it might still be
legitimate to explore the possibility that systematic experimental
errors be responsible for the paradox, and other solutions exist for
the GZK paradox) and we attempted to test the plausibility of a
description of the anomalies in terms of a Planck-scale-induced
deformation of Lorentz symmetry.  The results reported in Section~4
certainly indicate that this description is plausible. Had we not been
considering such a dramatic departure from conventional physics, we
would have probably gone as far as stating that the LIDs we considered
provide a compellingly simple description of the anomalies.  We do
feel that the results of Section~4, also taking into account that
there is no other known common explanation of the two threshold
paradoxes, provide motivation for additional theory work on the
speculative idea of LID and for additional experimental studies aimed
at testing the class of Planck-scale-induced LID here considered.
  
While presently-available data do not in any way invite one to look 
beyond the simplest two-parameter LID examined in Section~4, in 
preparation for future studies, especially the expected improvement of 
the experimental input, we have developed in Section~5 a general 
parameterization that may prove useful for future attempts to constrain 
(even rule out) Planck-scale-induced LID.  We have emphasized the fact 
that Planck-scale-induced LID, since it should reflect the 
structure of space-time, can be characterized by a small 
number of parameters.  In the high-energy regime we found that a very 
general description of (the leading effects of) Planck-scale-induced 
LID only requires five parameters and we described how the 
determination of a few thresholds together with measurements of the 
speed of very-high-energy particles could fix all five parameters.   
While we have considered a very general class of LID, it should 
be stressed that we postponed to future studies the analysis of  
an important class of further generalizations  
of Planck-scale-induced LID~\cite{gampul}: deformation terms involving
a dependence on polarization/spin of the particles.   
 
In closing we would also like to emphasize the fact that the 
experimental data here considered represent an important sign of 
maturity for the general programme of ``Planck-length 
phenomenology"~\cite{polonpap,gacPLph}.  Whether or not 
Planck-scale-induced LID turns out to successfully describe future 
experimental data, the fact that at present we are confronted with 
experimental paradoxes whose solution could plausibly involve the 
Planck length, and that certainly the relevant class of observations 
will eventually be able to rule out various pictures of the 
short-distance (possibly quantum) structure of space-time, shows that, 
contrary to popular folklore, some experimental guidance can be 
obtained for the search of theories capable of unifying gravitation 
and quantum mechanics.  This confirms the expectations, which were 
based on analyses of the sensitivity of various classes of 
experiments~\cite{ehns,grbgac,gacgwi}, that emerged from the general 
quantum-gravity studies reported in 
Refs.~\cite{polonpap,gacgwi,ahlunature,gacPLph} and 
from analogous studies, primarily focusing on the hypothesis that the 
unification of gravitation and quantum mechanics should involve 
non-critical strings, reported in Ref.~\cite{emnreview}. 
 
\bigskip 
\bigskip
\medskip
\noindent 
We thank Daniele Fargion and Glennys Farrar for many informative  
discussions on UHECRs.  
 
\baselineskip 12pt plus .5pt minus .5pt

\section*{Appendix A:  $\kappa$-Minkowski space-time} 
In order to illustrate in an explicit framework some of 
the structures relevant for our analysis of LID, in this Appendix  
we give a brief descrition of the ``$\kappa$-Minkowski" 
non-commutative space-time, which was developed in 
Refs.~\cite{gackpoinplb,kpoinap,gacmaj,lukipap,majrue}.
The simplicity of $\kappa$-Minkowski, which is basically an ordinary 
Minkowski space-time on which however one postulates that the time 
coordinate does not commute with the space  
coordinates ($[x_i,t] \sim x_i/E_p$),  
renders it very useful for the purpose of illustrating the 
new conceptual elements required by space-times with a nontrivial 
short-distance structure.   
 
A first point that deserves being 
emphasized is the connection between flat nontrivial space-times and 
quantum gravity.  In quantum gravity one has the general  
intuition~\cite{polonpap,gacgrf98} that ordinary classical 
commutative space-times should emerge from some more fundamental 
underlying picture. To very compact (Planckian-energy) probes 
space-time should look completely different from an ordinary classical 
space-times. On the contrary probes of very low energy should not be 
affected in any noticeable way by the nontrivial short-distance 
structure of space-time. In the intermediate regime (mid-energy 
probes~\cite{polonpap,gacgrf98}) 
one would expect to be able to use 
roughly the same language of ordinary classical space-times, but with 
the necessity to introduce some new concepts (such as the little 
element of noncommutativity of $\kappa$-Minkowski) reflecting the 
leading-order effects of quantum-gravity at low energies.  This 
hierarchy of regimes is to be expected not only in high-curvature 
space-times (where classical-gravity effects are stronger), but also 
in space-times that appear to be trivial and flat to  
very-low-energies probes.  $\kappa$-Minkowski is~\cite{gacgrf98}  
a model (toy model?) of how a probe of relatively high energy could  
perceive a space-time that instead appears to be trivially Minkowski 
to probes of very low energy. 
 
In $\kappa$-Minkowski a relation of type (\ref{dispone}) can be 
obtained as a direct consequence of the $\kappa$-Poincar\'{e} 
invariance~\cite{gackpoinplb,kpoinap,gacmaj,lukipap,majrue}
of this space-time. $\kappa$-Minkowski therefore provides an example 
of the mentioned scenario in which an ordinary symmetry 
is violated but there is no ``net loss of symmetries'' (the 
10-generator Poincar\'{e} symmetry is replaced by the 10-generator 
$\kappa$-Poincar\'{e} symmetry).  It is in order to capture the 
essence of these situations that one introduces the terminology 
``symmetry deformation" (in alternative to ``symmetry violation" 
which could be reserved for cases with a net overall loss of 
symmetries). In Sections~4 and 5 we denominate our scheme as a LID 
just to emphasize that the equations we use do not necessarily 
reflect a loss of symmetry (whether or not they do imply a  
net loss of symmetry depends on the underlying algebraic structures 
that lead to those equations in a given space-time picture). 
 
Importantly, consistency with the non-commutative 
nature of $\kappa$-Minkowski space-time also 
requires~\cite{gackpoinplb,kpoinap,gacmaj,lukipap,majrue} that the law
of addition of momenta be accordingly modified.  This modification 
emerges at the level of the $\kappa$-Poincar\'{e} (Hopf) algebra, and 
of course requires physical interpretation (particle momenta in a 
noncommutative space-time are a new concept).  A prescription suitable 
for handling the ambiguities due to the non-commutative 
nature of $\kappa$-Minkowski space-time was given recently in 
Ref.~\cite{gacmaj}, and in the cases here of interest, which always 
involve the sum of parallel momenta of two particles (at threshold 
particles are produced at rest in the CM frame), it 
reduces (in leading order in $E_{p}^{-1}$) to the prescription that 
the sum of momenta $K_1$ and $K_2$ can be handled with ordinary 
algebraic methods upon the  
replacement $K_1 + K_2 \rightarrow K_1 + K_2 + \delta K_1 K_2/ E_{p}$, 
where $\delta$ is a parameter analogous to $\eta$.  
In the analyses reported in Sections~4 and 5 this would  
imply $p_1 - \epsilon \rightarrow p_1 - \epsilon - \delta p_1 \epsilon/E_{p}$ 
and $p_2 + p_3 \rightarrow p_2 + p_3 + \delta p_2 p_3/ E_{p}$,  
and actually, since of course we have been here only 
interested in the leading $E_{p}^{-1}$ effect  
and $\epsilon \ll p_2 \sim p_3 \sim p_1$,  
one can neglect the term of order $p_1 \epsilon/ E_{p}$  
while retaining the term of order $p_2 p_3/ E_{p}$. 
 
\end{document}